\documentstyle[referee]{mn}
\input{epsf.sty}
\newif\ifAMStwofonts

\title{Formation of bulges in very late-type galaxies from SSCs}
\author[Y. N. Fu, J. H. Huang, and Z. G. Deng]
       {Y. N. Fu $^{1,4,5}$, J. H. Huang $^2$, and Z. G. Deng $^3$ \\
       $^1$ Purple Mountain Observatory, Chinese Academy of Sciences, Nanjing 210008, China\\
       $^2$ Department of Astronomy, Nanjing University, Nanjing 210093, China \\
       $^3$ Department of Physics, Graduate School, Chinese Academy of Sciences, Beijing 100039, China\\
       $^4$ Astronomie et Syst\`{e}mes Dynamiques, IMC-CNRS UMR8028, 77 Av Denfert-Rochereau, 75014 Paris, France\\
       $^5$ National Observatories, Chinese Academy of Sciences, Beijing 100039, China}

\date{Accepted .
      Received ;
      in original form}

\pagerange{\pageref{firstpage}--\pageref{lastpage}} \pubyear{}

\begin{document}

\maketitle

\label{firstpage}

\begin{abstract}

The dynamical evolution of super star clusters (SSCs) moving in
the background of a dark matter halo has been investigated as a
possible event responsible for the formation of bulges in 
late-type spirals. The underlying physical processes include  sinking
of SSCs due to the dynamical friction and  stripping of SSCs
on their way  to the center. 
Our model calculations show that only
sinking of circumnuclear SSCs contributes to the formation of 
galactic bulges at the early stage. Based on the assumption of a
universal density profile for the dark matter halo, and an 
isothermal model for the SSCs, our simulations have yielded 
bulges that are 
similar in many aspects to the observational ones. In particular,
the derived surface density profiles can be well fitted by an
exponential structure with nuclear cusps, 
which is consistent with HST observations.
\end{abstract}


\section{Introduction}

Bulges are basic building blocks of the Hubble sequence, as 
shown from the studies on the growth of galactic bulges by mergers
(e.g. Walker et al. 1996; Agueri et al. 2001), or by secular
evolution (e.g. Norman et al. 1996; Zhang 1999). It is important
to understand whether these processes could push
late-type spirals across the Hubble sequence toward early-type
spirals. Also, studies show that elliptical galaxies, which are 
thought to be similar to the bulges of early-type spirals, can 
be formed through mergers of spirals with similar
masses (e.g. Barnes \& Hernquist 1991).

A significant amount of  data on bulges have been accumulated (e.g.
Andredakis  et al. 1995; 
 Carollo 1999; Seigar et al. 2002), with a
common perception of $R^{1/4}$ and exponential bulges for early-
and late-type spirals, respectively. Among them, the HST survey of
early- to intermediate-type spirals (Carollo 1999, C99 hereafter) 
has yielded new
insight into the nuclear region of these systems.

The most interesting results from this survey are that 11 of the
12 S0a - Sab galaxies display $R^{1/4}$ bulges, while 14 out of
the 20 Sb - Sc galaxies have exponential bulges. The majority of
these bulges illustrate nuclear cusps, steeper ones for $R^{1/4}$
structures and shallower for exponential bulges. C99 argued
 convincingly that the shallower cusps in exponential bulges
are of stellar origin, and claimed that they are the old end of star
clusters with masses of about $10^6 M_{\odot}$.

Probably consistent with this argument is an evolved super star
cluster (SSC) located in the nucleus of M33 (Kormendy \& McClure
1993). The occurrence of old-end star clusters at the nuclei of
galaxies has been reported in other two nearby Sc galaxies, NGC
247 and NGC 2403 (Davidge \& Courteau 2002). In fact, nuclear star
clusters may be a common phenomenon (e.g. Davidge \& Courteau
2002; B\"{o}ker et al. 1999, and references therein) in late-type
galaxies. Possibly related to the existence of nuclear star
clusters is the presence of circumnuclear young massive star
clusters, or circumnuclear SSCs (e.g. Larsen \& Richtler 1999;
Whitmore et al. 1999;  Meylan 2001).

A question then arises whether there is a physical
relation between these two common phenomena in late-type spirals,
i.e., the common existence of nuclear/circumnuclear star clusters
and the exponential bulges with nuclear stellar cusps. If this is
the case, what is the physical process responsible for this
relation? A prevailing view for the formation of bulges in
late-type galaxies is the secular evolution of disks due to the
bar formation/destruction (e.g. Pfenniger \& Norman 1990; 
 Norman, Sellwood, \& Hasan 1996).
However, Bureau argued in his review (2002) against this scenario
based on the consideration of the fast duty cycle of the bar
formation/destruction, inferred from the omnipresence of bars
(Seigar \& James 1998).

In this paper we report our study, motivated by Carollo's claims
and the observations mentioned above, on the dynamical evolution
of circumnuclear SSCs moving in the dark matter halo as a possible
process for the formation of bulges with nuclear cusps in very
late-type galaxies.

It has been suggested (e.g., Tremaine et al. 1975) that
the dynamical friction acting on globular clusters
moving on orbits near the centre of a galaxy would lead to the
formation of a nucleus. The underlying physical processes
in our scenario, however, are sinking of SSCs
due to the dynamical friction and tidal stripping of SSCs on
their way to the center. The inclusion of tidal stripping,
which is not considered in the previous work, is crucial in
producing an exponential bulge with nuclear stellar cusp.

\section{Models}

\subsection{SSC model}
Recent high resolution observations of starburst galaxies
have revealed the existence of many compact, young, and very luminous
SSCs in the central regions of galaxies (Surace \& Sanders 1999, 
Scoville et al. 2000, de Grijs et al. 2001). Numerical simulations 
(Mihos \& Hernquist 1996, Barnes \& Herquist 1996) 
have also shown that interaction or merger among galaxies can
trigger strong starbursts around the nuclear regions. While
galaxies are merging or interacting, some gas components could
lose their angular momentum and fall into the central region. 
High pressure of the warm interstellar gas could induce global
collapse of giant molecular clouds and thus form circumnuclear
SSCs (Jog \& Solomon 1992, Harris \& Pudritz 1994).

Although there is already a body of observational data on SSCs,
their dynamical properties are not yet well understood.
SSCs are believed to be
the progenitors of present-day globular clusters (GC, e.g. 
 Origlia et al. 2001). It seems
reasonable to model SSCs, then, in light of this situation.

Based on the analogy with GCs, we assume that SSCs have a similar
mass spectrum as the initial GC mass function but with larger mean
value. According to Vesperini (2000,2001), the 
log-normal mass function for SSCs (SSCMF) can be written as
\begin{equation}
\label{SSCmassspectrum} Log_{10}(\frac{M}{M_{\odot}}) \sim  N(
\exp(mean)=2 \times 10^6 M_{\odot}, variance=0.08)
\end{equation}
where the mean value is chosen by the following consideration.
Model calculations indicate (e.g., Takahashi \& Portegies Zwart
2000) that compact SSCs with steep IMFs survive large mass loss
(more than 90\% of their initial masss) 
and thus are the most likely
progenitors of GCs. Considering the mean GC's mass of about $1-2
\times 10^5 M_{\odot}$, we may infer the mean SSC's mass of about
$2 \times 10^6 M_{\odot}$. 
The measured masses of SSCs detected in NGC4038/39 are consistent
with this prediction, being in the
range of $6.5 \times 10^5 - 4.7 \times 10^6 M_{\odot}$ (Mengel et
al. 2001), and 4 out of 5 clusters have masses larger than $2
\times 10^6 M_{\odot}$.

Using the above-mentioned SSCMF, we generate randomly 100 sets of
SSCs, each containing 100 SSCs (comparable to the number observed
by Fellhauer 2001 and Whitmore et al. 1999).
For simplicity, the SSC is modeled as a truncated isothermal
sphere with three parameters: the central density ($\rho_c$), the
velocity dispersion ($\sigma$), and the initial truncated radius
($R_0$). Taking a range  of $\rho_c$ between  $5.3
\times 10^3 M_{\odot}/pc^3$ and $3.4 \times 10^4 M_{\odot}/pc^3$
 (Larsen et al. 2001, Campbell et al. 1992), we
assume a linear function of cluster mass, $M$, for $\rho_c(M)$.
And with $R_0$ taken to be the local tidal radius, $\sigma$ can be
derived from $M$ and $\rho_c$.

According to the observational results of NGC1156, NGC1313, and
NGC5236 (Larsen \& Richtler 1999), the number of circumnuclear SSCs
is approximately proportional to $\frac{1}{R}$, where $R$ is the
projected distance from the galactic center. We assume that the
SSCs are distributed over the range of $0.4 - 6.4 kpc$ 
(Larsen \& Richtler 1999). All SSCs are
assigned a local circular speed.

\subsection{Background}

In general, small galaxies appear to be dominated by dark matter (DM)
even in their inner regions (e.g. Marchesini et al.
2002, Wilkinson et al. 2002, and references therein). In
particular, due to their bulge-less appearance and large
mass-to-light ratio, very late-type spirals are thought to be 
dark-matter-dominated at all radii. Analyses of high
resolution $H_{\alpha}$ rotation curves, in combination with
accurate HI rotation curves, confirms that late-type spirals
are completely dominated by dark matter (Persic et al. 1996,
Marchesini et al. 2002). Therefore, considering the
bulge-formation process in very late-type galaxies (Sdm type and
later) we only take a dark matter halo as the initial background. For
the dark halo, we assume the universal density profile (Navarro et
al. 1997, hereafter NFW), which can be written as (e.g., Binney \& Merrifield,
1998)
\begin{equation}
\label{NFWMassdensity} \rho_h(r)=\frac{M_{0h}}{r(a_h+r)^2}
\end{equation}
where $r$ is the distance from the halo center.

As the halos evolve , their
growth can have effect on the bulge formation process.  Here
we restrict our study to determining if the mechanism works
for SSCs formed in halos at different evolutionary stages.
We will focus on a time scale of 1 Gyr so that we can neglect
the evolution of the halo.  We consider two halos with quite
different masses:  a large one with mass
$M_{200}=10^{12} h^{-1} M_{\odot}$
(Y.P. Jing, private communication) and a smaller halo with mass
$M_{200}=3 \times 10^{11} M_{\odot}$. Further, we assume
a Hubble constant 
$H_{0} = 75 {\rm km s^{-1} Mpc^{-1}}$, $h$ = 0.75 
and a $\Lambda$CDM cosmological model (Jing, 2000).
Then the NFW halos are  determined via the scaling law  
(see Table 1).
\begin{center}
{\bf Table 1.} Parameters of the chosen NFW halos
\end{center}
\begin{center}
\begin{tabular}{c|c|c|c|c}
\hline
$M_{200} (M_{\odot})$ & $r_{200} (kpc)$ & $c$ & $a_{h}(kpc)$ & $M_{0h}(M_{\odot})$ \\
\hline
$10^{12}h^{-1}$ & 216.8 & 8.5 & 25.5 & $7.8 \times 10^{10}$ \\
\hline
$3 \times 10^{11}$ & 131.9 & 10 & 13.2 & $1.6 \times 10^{10}$ \\
\hline

\end{tabular}
\end{center}

While the mass outside several hundred $pc$s remains dominated by a dark
halo, the mass within  $100 pc$ can be
significantly changed by the fallen SSCs together with their
stripped mass. In order to account for this effect, 
we add a spherical component, represented by shells of equal
density, which varies according to the changing SSC mass
contribution.

\subsection{Dynamical friction}
We define  $M$ and $\vec{V}_M$ (with $V_M=|\vec{V}_M|$)
equal to, respectively, the mass and velocity of cluster
experiencing the dynamical friction. Assuming the background
matter has a Maxwellian
velocity distribution with dispersion $\sigma_{bkgd}$, and is
composed of particles with mass much smaller
than $M$, the dynamical friction formula may be written as
 (e.g., Binney \& Tremaine, 1987)
\begin{equation}
\label{Chdf} \frac{d\vec{V}_M}{dt}=-\frac{2 \pi \log(1+\Lambda^2)
G^2 M \rho}{V_M^3}[ {\rm erf} (X)-\frac{2X}{\sqrt{\pi}}
\exp(-X^2)] \vec{V}_M
\end{equation}
where {\rm erf} is the error function, and,
\begin{equation}
\label{quantiesinChdf} \left\{\begin{array}{l}
\Lambda=\frac{b_{max} V_{typ}^2}{GM} \\
X=\frac{V_M}{\sqrt2
\sigma_{bkgd}}
\end{array}\right.
\end{equation}

The quantity $b_{max}$ is the so-called maximum impact parameter
and $V_{typ}$ a typical measure of the relative velocity
between $M$ and a background particle. Neither $b_{max}$ nor
$V_{typ}$ is precisely defined. Fortunately, uncertainty in either
quantity causes no significant difference in the resulting value
of the dynamical friction. Following Binney  \& Tremaine (1987),
we use $b_{max} \equiv 2kpc$ and take $V_{typ} \equiv V_M$. The
velocity dispersion $\sigma_{bkgd}(r)$ can be roughly
estimated from the Jeans equation.

\subsection{Stripping}
We assume that the stellar mass outside a sphere of radius
$R_t$, which corresponds to the instantaneous Hill stability
region around the SSC center, will be stripped.  Since
stripping goes on continuously as $r$ (the distance between
halo and SSC centers) decreases, only a thin outer layer is 
stripped at a time. On average, the stars
stripped when the SSC goes
  from $r$ to $r-dr$ contribute to a
region radially bounded by $r+R_t(r)$ and $r-dr-R_t(r-dr)$. As a
first-order approximation, the mass of the stripped stars is
considered to be, at some later epoch, uniformly distributed in
the shell bounded by $r+R_t(r)$ and $r-dr-R_t(r-dr)$. By summing
up all of the stellar mass stripped at various $r$s, the stripped
stellar mass distribution can be derived.

If a single massive object is embedded in the center of an
SSC, stripping cannot proceeded further when only this object
is left. X-ray observations show many so-called
super-Eddington sources associated with SSCs (Matsumoto et al.
2001, Strikland et al. 2001). 
It is not yet clear whether the observations indicate intermediate-mass
black holes (IMBHs, ranging from several hundreds to about one thousand
solar masses, Ebisuzaki et al. 2001) or are due to the beam effect
(King et al. 2001).
If massive black holes do form in SSCs,
they would most likely be at the center. Therefore, in
our simulations, we consider two extreme cases:  stripping is
not allowed when the mass of the stripped SSC is less than $1
M_{\odot}$ and $1000M_{\odot}$, respectively.

\section{Results}

\begin{figure*}
{\epsfxsize= 15.5 cm \epsfbox{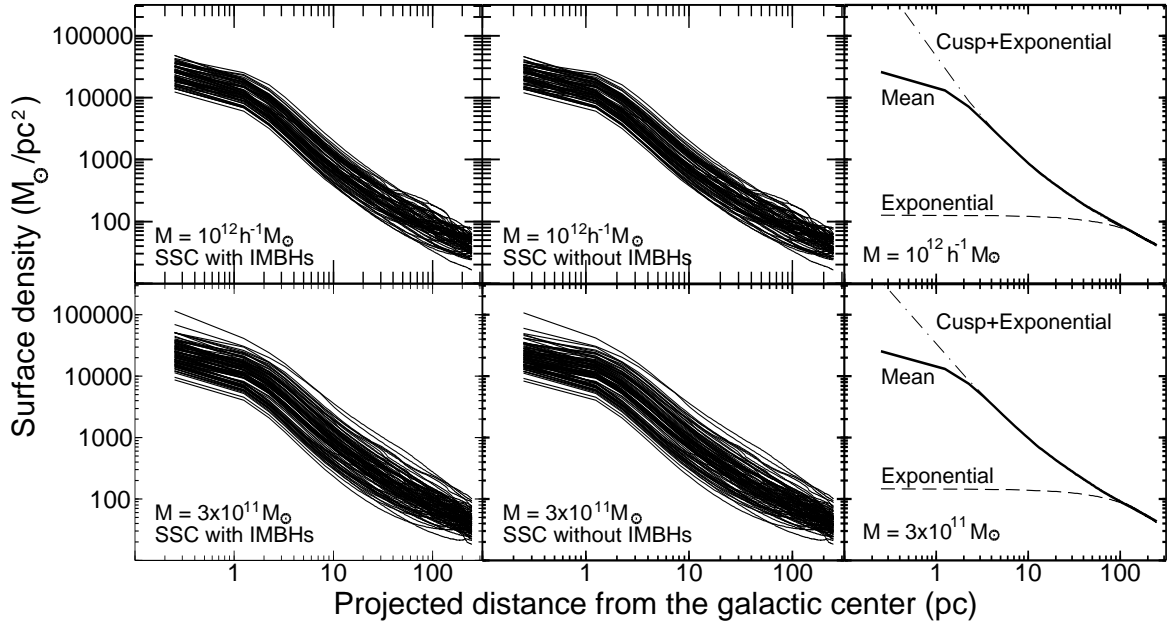}}\caption{Surface densities
derived by SSCs with/without IMBHs for two DM density profiles adopted.
 The mean surface density profiles illustrated
in the right two panels are derived from the models without IMBHs. }
\end{figure*}
Figure 1 shows the surface density profiles of simulated individual
galaxies with high-density central components formed at 1 Gyr (with a 
proportion of about 65\% and 90\% for the larger and smaller
halos, respectively). Comparing our
simulations with the results of the HST survey by Carollo et al. (1999)
 is very instructive. Among their surveyed sources,
13 galaxies are of Hubble types Scd and later.
For these galaxies (at distances greater than 13 Mpc)
the smallest area the HST can resolve is  5 pc from 
the center.
About 50\% of the 13 galaxies contain exponential bulges with
central stellar clusters. The remaining ones do not have accurate
fits, though some of them display central cusps. 
Our mean model profiles (also in Fig.1) share some
characteristics with the above observational results, i.e., the general
presence of resolved compact clusters
on top of the exponential bulges.

In order to give a quantitative comparison with the
observations, following Carollo et al., we first fit the outer part of
the mean profile with an exponential model.  We then subtract this model
of the exponential bulge, leaving the cusp component. This, in turn, is
fit with the cusp model of Carollo \& Stiavelli (1998, eq(3)),
which was also used to fit the bulge of one Sd galaxy (ESO499 G37). The
overall model is thus given by
\begin{equation}
\label{model} \sigma(R)=\sigma_0 \exp(-1.678
\frac{R}{R_e})+\sigma_1(1+\frac{R_c}{R})^\gamma
\exp(-\frac{R}{R_s})
\end{equation}
The resulting profiles are also shown in Fig. 1, and the associated
values of the parameters are given in Table 2.
Except at radii less than about 2 pc, the fittings shown in Fig. 1
are quite satisfactory and
are in agreement with HST observations.
Recent HST observations, however, can not 
discriminate such a difference within about 2 pc, as we described above.

\begin{center}
{\bf Table 2.} The fitted values of parameters in (\ref{model})
\end{center}
\begin{center}
\begin{tabular}{c|c|c|c|c|c|c}
\hline $M_{200}$ & $\sigma_0$ & $R_e$ & $\sigma_1$ & $R_c$ & $\gamma$ & $R_s$ \\
       $(M_{\odot})$ & ($M_{\odot}/pc^2$) & (pc) & ($M_{\odot}/pc^2$) & ($pc$) &  & ($pc$) \\
\hline $10^{12} h^{-1}$ & 127 & 361 & 317 & 8 & 2.2 & 23 \\
\hline $3 \times 10^{11}$ & 146 & 327 & 212 & 18 & 1.7 & 27 \\
\hline
\end{tabular}
\end{center}
The consistency between observations and simulations
indicates that the dynamical evolution of SSCs proposed in this
paper could be an explanation for the formation of exponential bulges
with central stellar cusps for very late-type galaxies.

\begin{figure}
{\epsfxsize= 9 cm \epsfbox{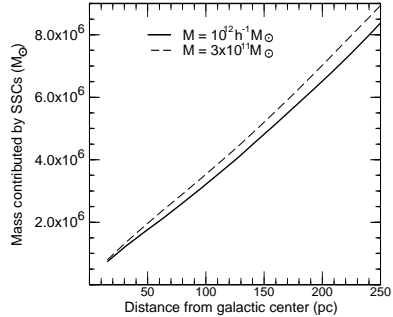}} \vspace{-4cm} \caption{
Mean mass profiles contributed by SSCs at 1 Gyr, showing different stripping
 effects  of two DM density profiles adopted over bulge area}

\end{figure}

\begin{figure}
{\epsfxsize= 12 cm \epsfbox{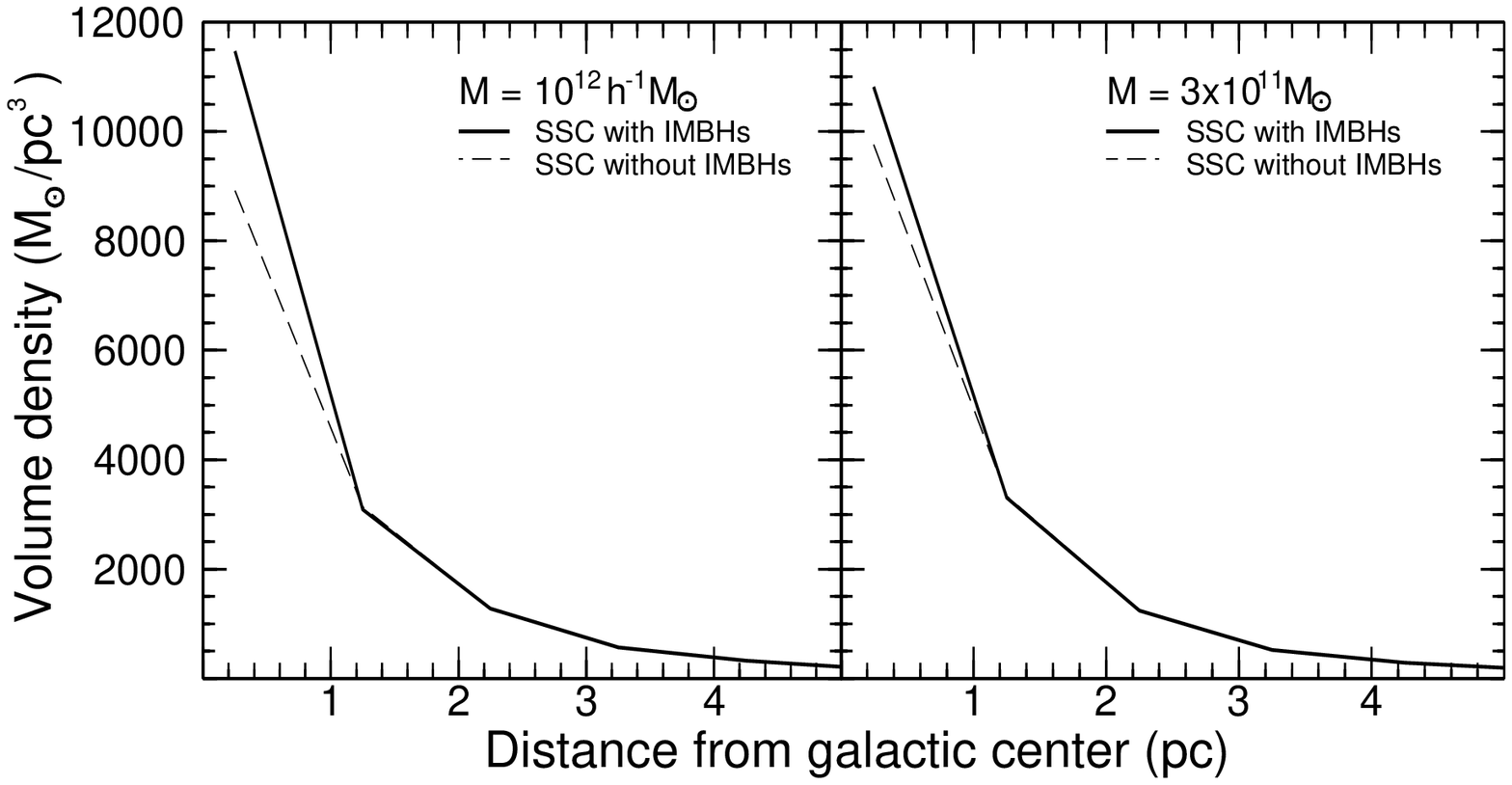}} 
\vspace{-0.5cm} \caption{Volume densities contributed by SSCs with/without 
IMBHs for two adopted DM density profiles, indicating the diferences of formed
bulges at about 1 pc.  }
\end{figure}

As mentioned above, the proportion of the bulges formed at 1 Gyr
in the larger halo is significantly less than in the smaller halo,
65\% vs.\  90\% .
This is because larger halos have a stronger stripping effect on
SSCs at a large distance from the center (due to their less centrally
concentrated mass distribution), and, therefore, SSCs become less massive
there, causing slower sinking to the bulge area. In a
smaller halo, the SSCs can deposit more of their mass in the inner
regions, as shown in Fig.2, resulting in  more significant cusp components
with $R_c$ a factor of 2 larger than in  the larger halo, 
as indicated in Table 2.
The close correspondences between other parameters in Table 2
follow from the similarities of the NFS density profiles with 
different masses (Navarro et al. 1997).

In order to see what could happen to the galaxies
where no high-density central component is formed at 1 Gyr, we
extend the simulation to 1.5 Gyr. The bulge formation rates are
then increased to about 90\% and 99\% for the larger and smaller
halos, respectively. This result implies that the proposed
mechanism may be a viable mode  for the formation of bulges and
the nuclear star clusters.

The properties of the formed bulges, as shown in Fig. 1,
are largely independent of the existence of IMBHs in SSCs, except
for the region very near the galactic center (with radius of about
1 pc), as shown in Fig. 3. The difference arises because the
tidal disruption cannot strip the SSC with only an IMBH left.

\section{Discussion}

In our simulation, we have made some assumptions either due to the
lack of information or for simplicity. An important
assumption is that the background is composed of dark matter only,
though we have considered the background variation later. Indeed,
this assumption is compatible with what we investigate in this
paper: the formation of bulges at the early stage. 
Accordingly, the SSCs we have studied are circumnuclear
ones presumably originating from the mergers of very late-type
galaxies. These galaxies have disk components only. In
this case, the dark matter dominates over luminous systems at all
radii.
As a first step of our investigations it is reasonable to
make such an assumption. On the other hand, it is interesting to
see what the theoretical prediction will be if more components are
added, such as disks and a pre-existing bulge, in the background.
The logical next step of our
investigations is to study a way for the first-formed bulge to
grow up further. Probably a new merger is needed. A new
merger occurs between two disk galaxies with small bulges. A study
of this kind is under our consideration.


As mentioned in Sect. 3, different halos will produce
different bulge formation rates (more concentrated halos tend to be
more effective in producing bulges). To try other halo models is
under our consideration. Also, it would be very intriguing to
consider a more realistic case, i.e., the one of halo variation
incorporating several starbursts.

There would certainly be accumulated compact objects from the
stripped inner part of the SSCs, as well as the remaining SSCs
themselves---such as neutron stars, stellar BHs, or LMXBs with BHs---in
the deep potential well of the galactic center. Whether these
objects can form a BH at the galactic center is worth further study.

\section*{Acknowledgments}

The authors would like to thank Dr.\ Mark Wilkinson, the referee,
for his constructive comments, which helped to
substantially improve the manuscript. Drs.\ W. Zheng and R. Hanisch 
are thanked for their kind 
help in improving the English.
The authors are also
grateful to Drs.\  Y.P. Jing, X.P. Wu, X.Y. Xia and C.G. Shu. for their valuable
discussions. Fu is greatly indebted to the colleagues in
Astronomie et Systemes Dynamiques, IMC-CNRS UMR8028. This work is
supported by NKBRSF G19990754
and NSFC.

\label{lastpage}

\end{document}